\documentclass[preprint,preprintnumbers,aps,prb]{revtex4}
\usepackage{graphicx}
\usepackage{amssymb}

\def\bef2{BeF$_2$} 
\def\sio2{SiO$_2$} 
\def\jmatc#1#2#3{{ J. Mater. Chem.} {\bf #1}, #2 (#3).}

\def\molsim#1#2#3{{ Mol. Simulat.} {\bf #1}, #2 (#3).}

\def\physicaa#1#2#3{{ Physica A} {\bf #1}, #2 (#3).}
\def\physicab#1#2#3{{ Physica B} {\bf #1}, #2 (#3).}

\def\prl#1#2#3{{ Phys. Rev. Lett.} {\bf #1}, #2 (#3).}

\def\pre#1#2#3{{  Phys. Rev. E.} {\bf #1}, #2 (#3).}
\def\pra#1#2#3{{ Phys. Rev. A} {\bf #1}, #2 (#3).}
\def\prb#1#2#3{{ Phys. Rev. B} {\bf #1}, #2 (#3).}
\def\jcp#1#2#3{{ J. Chem. Phys.} {\bf #1}, #2 (#3).}
\def\jpc#1#2#3{{ J. Phys. Chem} {\bf #1}, #2 (#3).}

\def\jpcb#1#2#3{{ J. Phys. Chem. B} {\bf #1}, #2 (#3).}
\def\jsp#1#2#3{{ J. Stat. Phys.} {\bf #1}, #2 (#3).}
\def\cpl#1#2#3{{ Chem. Phys. Lett.} {\bf #1}, #2 (#3).}
\def\cp#1#2#3{{ Chem. Phys.} {\bf #1}, #2 (#3).}

\def\mp#1#2#3{ { Mol. Phys.} {\bf #1}, #2 (#3).}

\def\science#1#2#3{{ Science} {\bf #1}, #2 (#3).}
\def\nature#1#2#3{ { Nature} {\bf #1}, #2 (#3).}

\def\jpcm#1#2#3{ { J. Phys. Cond. Matt.} {\bf #1}, #2 (#3).}

\def\pccp#1#2#3{ { Phys. Chem. Chem. Phys.} {\bf #1}, #2 (#3).}
\def\jncs#1#2#3{ { J. Non-Cryst. Solids} {\bf #1}, #2 (#3).}

\def\be{\begin{equation}}
\def\ee{\end{equation}}

\def\expt#1{\langle #1\rangle}

\def\br{{\bf r}}

\begin{document}
\baselineskip 20pt
\begin{center}
{\Large \bf Estimating Entropy of Liquids from Atom-Atom Radial
Distribution Functions: Silica, Beryllium Fluoride and Water}\\
\qquad \\
{\bf Ruchi Sharma}, {\bf Manish Agarwal} and
{\bf Charusita  Chakravarty}$^*$\\
Department of Chemistry,\\
Indian Institute of Technology-Delhi,\\
New Delhi: 110016, India.\\
 \quad\\
 {\bf Abstract}
 \end{center}

 Molecular dynamics simulations of water, liquid beryllium fluoride and 
 silica melt are used to study the accuracy with which the  entropy of
 ionic and molecular liquids can be estimated from atom-atom
 radial distribution function data.
The pair correlation entropy is demonstrated to be sufficiently
accurate that the density-temperature regime of anomalous behaviour
as well as the strength of the entropy anomaly 
can be predicted reliably for both ionic melts as well as different
rigid-body pair potentials for water.  Errors in the total
thermodynamic entropy for ionic melts  due to the pair
correlation approximation are of the order of 10\% or less for most state points
but can be significantly larger in the anomalous regime at very low 
temperatures.  In the case of water,  the rigid-body
constraints result in  larger errors in the pair correlation approximation,
between 20 and 30\%, for most state points.
Comparison of the excess entropy, $S_e$, of  ionic melts with
the pair correlation entropy, $S_2$, shows that the temperature dependence of 
$S_e$ is well described by $T^{-2/5}$ scaling across both the normal and 
anomalous regimes, unlike in the case of $S_2$.  
The residual multiparticle entropy, $\Delta S=
S_e-S_2$, shows a strong negative correlation with tetrahedral order in the 
anomalous regime.
	
\vfill
{* Author for correspondence (Tel: (+)~91~11~2659~1510; Fax: (+)~91~11~2686~2122; E-mail: {\tt charus@chemistry.iitd.ernet.in})\hfill}
\newpage

\section{Introduction}

The thermodynamic excess entropy of a fluid is defined as the difference in 
entropy between the fluid and the corresponding ideal gas under identical 
temperature and density conditions. In the case of a classical fluid, 
the excess entropy corresponds to the lowering of the entropy of the fluid, 
relative to the ideal gas, due to the presence of multi-particle positional 
correlations \cite{hm86}. The total entropy of a classical fluid 
can  be written as \cite{hsg52,ng58,hjr71,dcw87,be89,be90,bb92,ggg92} 
\be
S=S_{id} + S_2 + S_3 + \dots = S_{id} +\sum_{n=2}^\infty S_n
\ee
where $S_{id}$ is the entropy of the ideal gas reference state,  $S_n$
is the entropy contribution due to $n$-particle spatial correlations and
the excess entropy is defined as, $S_e=S-S_{id}$. 
Such a multiparticle expansion  of the entropy is interesting because it 
allows for the prediction
of thermodynamic properties from structural correlation functions.
Moreover,  semiquantitative excess-entropy based  scaling relationships of 
the form 
\be X=A\exp (-\alpha S_e)\ee
 where $X$ is a suitably scaled transport property 
 \cite{yr77,yr99,md96,yr00,has00} allow for the possibility of making
 interesting connections between structural, thermodynamic and transport
 properties of fluids.

Neutron or X-ray scattering experiments can provide atom-atom
radial distribution functions (RDFs), $g_{\alpha\beta}(r)$, associated with the
probability of finding an atom of species $\beta$ at a distance $r$ from
an atom of species $\alpha$ relative to the probability in 
the corresponding ideal gas \cite{hm86}. In terms of the atom-atom radial 
distribution function, one can write an ensemble-invariant expression for the 
pair correlation contribution to the entropy of a multicomponent fluid of
$N$ particles enclosed in a volume $V$ at temperature $T$ as\cite{jah90,lh92}
\be
S_2/Nk_B = -2\pi\rho\sum_{\alpha ,\beta } x_{\alpha }x_{\beta }
\int_0^\infty 
\{g_{\alpha\beta}(r)\ln g_{\alpha\beta}(r) -\left[ g_{\alpha\beta}(r)-1\right] \} r^2 dr 
\ee
where $x_\alpha$ is the mole fraction of component $\alpha$ in the mixture
and $\rho$ is the number density of the fluid.
Note that the excess entropy is measured in this case with respect to
the entropy of an ideal gas consisting of a non-interacting mixture of 
spherical particles with entropy $S_{id}$ given by
\be
S_{id}/Nk_B = 1 + \sum_\alpha x_{\alpha } \left[0.5v_\alpha -\ln (\rho_\alpha
\Lambda_\alpha^3 )\right]
\ee
where $v_\alpha$ is the number of degrees of freedom associated with the 
component $\alpha$ and $\Lambda_\alpha$ is the thermal de Broglie wavelength of 
component $\alpha$. 
The multiparticle expansion of the entropy  has been studied most extensively
for simple liquids and liquid mixtures  of structureless particles with 
isotropic interactions \cite{hjr71,dcw87,be89,be90,bb92}. For such systems, 
the pair correlation contribution captures 
85-90\% of the excess entropy, with the largest discrepancy occurring at 
intermediate densities \cite{bb92}. 

{ In the case of a homogeneous, isotropic molecular fluids, 
an alternative and physically reasonable reference state for the multiparticle
expansion is provided by an ideal gas of rigid rotors. In this case,
pair correlation contribution to the entropy must be formulated in terms
of orientationally-dependent pair correlation functions, $g(r,\omega^2)$, where
$\omega^2$ denote the relative orientations of the two molecules 
\cite{lp92,lp94}. 
The orientation-dependent pair correlation
It is convenient to perform a further decomposition of
$g(r,\omega^2)$ as
\be g(r, \omega^2) = g(r) g(\omega^2|r)\ee
where $g(\omega^2\vert r)$ is the conditional probability of observing an
orientational separation $\omega^2$ at a relative separation $r$ of the
molecular centres of mass.
The pair correlation entropy of a molecular
system, denoted by $S_2^{mol}$, may then be written as a sum of
the translational entropy, $S_2^{tr}$,  dependent only on $g(r)$,
and orientational contribution, $S_2^{or}$, involving a suitable
weighted average over $g(\omega^2 |r)$. While simulations show
that  this procedure yields reasonable estimates of the entropy for standard 
pair-additive water models,  orientation-dependent pair correlation
functions are not directly accessible from experiment. Since evaluation of the
correlation functions from simulations with adequate statistical accuracy
may prove complicated, a number of approximate factorizations of
$g(r,\omega^2)$ have been studied \cite{lk96,jz05,jz08}.}

Since  atom-atom radial distribution functions (RDFs) are obtainable
from experiments, obtaining estimates of thermodynamic
and transport properties of liquids from atom-atom RDFs is
very attractive because it allows one to  directly connect structural
information with thermodynamic and transport properties
of fluids. In principle, such studies could contribute to improving reverse
Monte Carlo strategies for structure determination as well as 
multi-scale methodologies for constructing coarse-grained potentials
\cite{km90,bl07,aal02,jhl07,hs93,lh06,mrytd07}.
In order to develop this possibility, it
is necessary to test the numerical validity of the pair correlation
approximation for a wider range of systems than simple atomic liquids.
The purpose of this paper is to examine the extent to which the pair
correlation entropy, computed from atom-atom RDFs, captures the
temperature and density dependence of the excess entropy for
ionic melts and molecular liquids.  We also study the behaviour of
the residual multiparticle entropy (RMPE),
defined as $\Delta S=S_e-S_2$,  since it has been shown contain physically
significant information on local order in a liquid close to 
phase transitions. { For example, the state point 
for which the RMPE crosses from negative to positive is
found to be close to ordering phase transitions, such as freezing  and the 
isotropic-nematic
transition \cite{ggg92,yr00,csg03,ssg03}. }

Ionic melts are of considerable
technological importance and are, as a consequence, 
well-studied experimentally as well as computationally
\cite{mw00,pfm04}.  Many ionic melts form random liquid state networks and 
show interesting deviations from simple liquid behaviour in their
thermodynamic and transport properties \cite{pha97,ms98,abhst00,ht02,trcp07}.
Other than a limited comparison in the case of
liquid silica \cite{scc06,asc07,ac07}, we are not aware of any systematic 
testing of the pair correlation approximation  for the entropy in the
case of ionic melts.  As representative examples
of ionic melts, we choose beryllium fluoride (BeF$_2$) and silica (SiO$_2$).
Molecular liquids represent an interesting test
case for estimating entropy using atom-atom RDF data,
since the pair correlation approximation, as defined in equation (3), is
expected to be poorer because the bond angle contraints give rise to strong 
three-body correlations  and an ideal gas of rigid molecules is a more
appropriate reference state, than a mixture of monoatomic
ideal gases as implicit in equation (3).  
Nonetheless, the experimentally accessible quantities are the atom-atom
RDFs and it is interesting to see how much information can be extracted
from these quantities.
As an example of a molecular liquid, we choose water which is of obvious 
interest as a pure liquid as well as a solvent, and is representative
of a wider class of hydrogen-bonded solvents \cite{sbr97,sr00,aks00,sdfc06}.
Computing the entropy of a molecular fluid
from atom-atom pair correlation data has not been attempted, though
orientationally dependent pair correlation 
functions, which are not accessible  experimentally,
have been used to estimate the entropy of several commonly used
model water potentials \cite{jz05,essg06}.

All three liquids that we study in this work (H$_2$O, SiO$_2$ and BeF$_2$) have 
similar anomalous thermodynamic and kinetic properties
\cite{scc06,asc07,ac07,pha97,ms98,abhst00,ht02,trcp07,sps01,ssp04,ssp00,sdp02,hma01,sslss00,ed01}. The most 
obvious signature of the thermodynamic anomalies is the existence of a 
regime of anomalous density behaviour where the isothermal expansion 
coefficient, $\alpha =(1/V_m)(\partial V_m/\partial T)_P$ is negative.
The region of the density anomaly is
bounded by temperatures of maximum and minimum density for which
$\alpha = 0$. While the temperature of minimum  density is experimentally
difficult to observe, the locus of temperatures of maximum density (TMD)
can be traced in the density-temperature or temperature-pressure plane
for many systems. The maximum temperature and maximum density along this locus, $T_{TMD}^{max}$ and $\rho_{TMD}^{max}$ respectively, can be thought of as upper
bounds above which the thermal kinetic energy or degree of compression 
respectively
is sufficient that the liquid conforms to thermodynamic behaviour characteristic
of simple liquids. Note that the lowest temperature or maximum density
that can be obtained on the TMD locus will be limited by spontaneous
crystallisation or glass formation.
The diffusional anomaly corresponds to the set of
state points for which the diffusivity  increases with density.
A unified explanation for the thermodynamic and kinetic anomalies can
be found based by existence of an excess entropy anomaly involving a
rise in excess entropy on isothermal compression 
\cite{etm06,scc06,asc07,ac07,met061,met062}.  The relationship between excess 
and pair correlation entropies in these liquids is of special interest from 
the point of view of  comparing the two quantities in the normal and 
anomalous regimes, since there is some indication that anomalous colloidal 
fluids with core-softened interactions show a greater magnitude of the RMPE 
contribution at  high densities \cite{etm06}.  { In the case of water,
simulation results for TIP4P water suggest that the
RMPE is close to zero close to the TMD at 1 atm pressure.}

Molecular dynamics simulations for BeF$_2$ and SiO$_2$ are performed using
the transferable rigid ion model (TRIM)\cite{ha97,hma01,wac76,wrb72} and 
van Beest-Kramer-van Santen (BKS)\cite{bks90,kfb91,ssp00}
potentials respectively. For both the systems, the ideal gas limit
is defined as an MX$_2$ binary mixture of uncharged particles with the same
masses as the corresponding ionic melts but with zero interparticle
interactions. The excess entropy, $S_e$, with respect to this ideal gas is
evaluated using thermodynamic integration. Using the radial distribution
functions extracted from the simulations, one can obtain both the pair
correlation entropy ($S_2$) and the residual multiparticle entropy (RMPE), 
given by $\Delta S= S_e - S_2$.  In the case of water, we have
simulated three of the commonly used rigid-body, effective pair
interaction models(SPC/E\cite{bgs87}, TIP3P\cite{jcmik83} and TIP5P\cite{mj00}) 
and  evaluated  the entropy
from the $g_{OO}(r)$, $g_{HH}(r)$ and $g_{OH}(r)$ radial distribution
functions. Table I shows the $T_{TMD}^{max}$ and $\rho_{TMD}^{max}$ values
for BKS silica, TRIM BeF$_2$ and the SPC/E and TIP5P water models
as determined in previous studies \cite{asc07,sdp02,ed01,sbgn03,ymss02}.

The paper is organised as folllows. Section II summarises the
computational details associated with the molecular dynamics simulations
of silica, beryllium fluoride and water that we have performed. 
Section III describes the thermodynamic integration
procedure that we have followed to obtain the thermodynamic
entropy of the two MX$_2$ ionic melts. In the case of water,
thermodynamic entropies for standard potentials are available
in the literature. Section IV discusses the results for the two ionic
melts while Section V contains the results for water. Conclusions
are summarised in Section VI.

\section{Computational Details}

For all the three systems, we use parametric effective pair potentials.
Each of these interaction models has been well-studied in the literature
in terms of behaviour on supercooling, glass transition and 
waterlike structural, kinetic and thermodynamic anomalies.
For convenience, we describe the functional forms of the pair potentials
and give the parameters in Tables II, III and IV. The molecular dynamics
simulations were performed using the DL\_POLY software package 
\cite{sf96,syr01}.

\subsubsection{Model Potential for Silica}

To model the interatomic interactions in silica, we use the van 
Beest-Kramer-van Santen (BKS) potential with an additional 30-6 Lennard-Jones
type correction term \cite{bks90,kfb91,ssp00}. 
The pair interaction between atoms $i$ and $j$ is given by:
\begin{equation}
\phi_{BKS}(r_{ij}) = \frac{q_iq_j}{4\pi\varepsilon_0r_{ij}} +
A_{ij}exp^{-b_{ij}r_{ij}} - \frac{C_{ij}}{r^6_{ij}}
+ 4\epsilon_{ij}\left[\left(\frac{\sigma_{ij}}{r_{ij}}\right)^{30} - 
\left(\frac{\sigma_{ij}}{r_{ij}}\right)^6\right]
\end{equation}
where $r_{ij}$ is the distance between atoms $i$ and $j$ carrying charges
$q_i$ and $q_j$,  $A_{ij}$, $b_{ij}$ and $C_{ij}$ are the parameters
associated with the Buckingham potential for short-range repulsion-dispersion
interactions and $\epsilon_{ij}$ and $\sigma_{ij}$ are the energy and length 
scale parameters for the 30-6 Lennard-Jones interaction. The parameters for the
modified BKS potential used in this work are given in Table II.

\subsubsection{Model Potential for Beryllium fluoride}

We use the transferable rigid ion model (TRIM) potential  for interatomic 
interactions  in beryllium fluoride \cite{wac76,wrb72,hma01,ha97}. 
The pair interaction between atoms $i$ and $j$ is given by:
\be
\label{eqn:phitrim}
\phi_{TRIM}(r_{ij}) = \frac{z_iz_je^2}{4\pi\varepsilon_or_{ij}} + 
\left(1 + \frac{z_i}{n_i} + \frac{z_j}{n_j} \right)
b\exp{\left(\frac{\sigma_i + \sigma_j - r_{ij}}{s}\right)}
\ee
where $r_{ij}$ is the distance between atoms $i$ and $j$. The parameters
associated with an atom of type $l$ are the charge $z_l$, the number of 
valence-shell electrons $n_l$ and the ionic size parameter  $\sigma_l$. 
The repulsion parameter $b$ and the softness parameter $s$ are assumed 
to be the same for all three types of pair interactions in BeF$_2$. 
The parameters for the TRIM potential used in 
this work are given in Table III. 

\subsubsection{Model Potentials for Water}

All rigid-body, effective pair potentials for water assume that
the molecule can be represented by a single Lennard-Jones site
located on the oxygen atoms and model the charge distribution
by a set of distributed charges with a fixed geometry.
The parametric form of the interaction between
two water molecules $a$ and $b$ is given by:
\be
U_{ab} = \sum_i\sum_j\frac{q_{i} q_{j}}{r_{ij}} + 4\epsilon\left(
\frac{\sigma^{12}}{r_{OO}^{12}} - \frac{\sigma^6}{r_{OO}^6}\right)
\ee
where $i$ and $j$ index partial charges located on molecules $a$ and $b$ respectively
and $r_{OO}$ refers to the distance between the oxygen atoms of the two
monomers.

The SPC/E and TIP3P models use 
 three charged sites which are located at the 
atomic positions and are associated with the corresponding atomic masses. 
The TIP5P model retains the three atomic sites and uses two additional
massless sites to represent the oxygen lone pair electron density distribution.
The potential parameters for the three water models are summarised in
Table IV.

\subsubsection{Molecular Dynamics}

Molecular Dynamics simulations for all three systems were 
carried out in the canonical ({\it N-V-T}) ensemble, using the DL\_POLY
software package \cite{sf96,syr01}, under cubic periodic boundary conditions.
The effects of electrostatic (long-range) interactions were accounted for by
the Ewald summation method \cite{at86,fs02}. The non-coulombic part was
 truncated and shifted at 7.5${\mathring{A}}$ for SiO$_2$ and BeF$_2$
 and at 9.0${\mathring{A}}$ for H$_2$O.
 A Berendsen thermostat, was used to maintain the
desired temperature for the production run. The leapfrog Verlet algorithm
with a time step of 1fs was used to integrate the equations
of motion; in the case of water, the SHAKE algorithm was used to impose
the rigid-body contraints. The MD simulation details, including the
lengths of the equilibration and production runs, are summarised in
Table V. The cutoff distance for the radial distribution functions
was kept the same as the potential cutoff and the bin size was
adjusted so that trapezoidal qaudrature to evaluate the
pair correlation contribution to the entropy using equation (3)
resulted in less than 1\% error.

\section{Estimation of Thermodynamic Entropy of Ionic Melts}

We follow the thermodynamic integration procedure described previously
by Saikia-Voivod and co-workers to evaluate the thermodynamic excess
entropy of BKS silica where the full Ewald summation was replaced by 
smoothly tapering off to zero the real-space and short-range contributions 
to the interaction and removing the reciprocal space
contribution\cite{sslss00,sps01,ssp04,ssp00}. The internal energy
differences between the short-range and full Ewald versions of the
BKS were of the order of 0.2\% or less. Since we use the full Ewald 
version of the potential, it was, however, necessary to recompute
the thermodynamic entropy of BKS silica. To our knowledge, the
entropy of the TRIM model of BeF$_2$ has so far not been evaluated.

The  model potentials  used here
envisage the ionic melts as being fluids composed of particles with
long-range Coulomb interactions and short-range repulsions.
Such a Coulombic
system cannot be reversibly transformed into an ideal gas reference system.
Therefore, a binary Lennnard-Jones (BLJ) system with an MX$_2$ stoichiometry
was introduced as an intermediate state. 
The entropy of the ionic melt at a given state point $(V_0,T_0)$ was
first evaluated relative to the binary Lennard-Jones system at the
same state point. The entropy of the  BLJ liquid at $(V_0,T_0)$ was 
then evaluated relative to the ideal gas limit at $(V_\infty, T_0)$ where
$V_\infty$ corresponds to such a large volume that the fluid can be
regarded as an ideal gas.

To determine the entropy of the ionic melt relative to the BLJ liquid, 
thermodynamic integration was performed with respect to a Kirkwood-type coupling
parameter, $\lambda$,  which interpolates between the potential energy function,
$U(\br )$, of the BLJ liquid and the ionic melt such that\cite{fs02}
\begin{equation}
U_{\lambda}(\br )=(1- \lambda) U_{BLJ}(\br )+\lambda U_{MX_2}(\br )
\end{equation}
The free energy difference between the BLJ and $MX_2$ ionic melt at $(V_0,T_0)$ 
can be written as:
\begin{equation}
\Delta F(V_0,T_0) = \int ^1_0 \left \langle \frac {\partial U_{\lambda}}{\partial \lambda} \right \rangle _{\lambda} d\lambda = 
\int ^1_0 \langle U_{MX_2}- U_{BLJ}\rangle_{\lambda} d\lambda
\end{equation}
The integrand $\langle\cdots\rangle_\lambda $ in the above equation refers
to a canonical ensemble average for a system described by the
potential energy function $U_\lambda (\br )$.
The integral   was evaluated using trapezoidal quadrature with the value
of the integrand at specific values of the variable $\lambda$ being
computed using  a canonical ensemble MD simulation with volume $V_0$,
temperature $T_0$ and interaction potential $U_\lambda (\br )$.
The entropy of system $MX_2$ at this state point, denoted by 
$S_{MX_2}(V_0,T_0)$, then be evaluated using,
\begin{equation}
S_{MX_2}(V_0,T_0) = S_{BLJ}(V_0,T_0) + \frac{1}{T} \left[ U_{MX_2}(V_0,T_0) - U_{BLJ}(V_0,T_0) - \Delta F(V_0,T_0)\right]
\end{equation}

The entropy of the BLJ system relative to the binary ideal gas is obtained as
\begin{equation}
S_{BLJ}(V_0,T_0)=S_{id}(V_0,T_0)+\frac {1}{T}\left [
\expt{U_{BLJ}(V_0,T_0)}-\int ^{V_\infty}_{V_0} P^{ex}_{LJ} dV\right ]
\end{equation}
where $S_{id}$ is the entropy of the ideal gas as defined in equation (4)
The third term in the above expression is obtained by
evaluating the excess pressure of the BLJ system for increasing system
volumes using simulations at moderate to low densities and a virial expansion
at very low densities. 

Once the entropy of the ionic melt has been obtained at a given state
point $(V_0, T_0)$ using the above procedure, the entropy at an
arbitrary state point $(V,T)$ can be obtained as
\begin{equation}
S(V,T) = S(V_0,T_0) + \Delta S_T + \Delta S_V
\end{equation}
where $\Delta S_T$ is  entropy change for isothermal
change in volume from $V_0$ to $V$ and $\Delta S_V$ is the entropy change for
isochoric change in temperature from $T_0$ to $T$. The following
expression can be used to evaluate $\Delta S_T$:
\begin{equation}
\Delta S_T = \frac{1}{T_0} \left[ U_{MX_2}(V,T_0) - U_{MX_2}(V_0,T_0) + \int^V_{V_0} P(V')dV'\right]
\end{equation}
where $P$ is the pressure of the MX$_2$ liquid.  To evaluate $\Delta S_V$, 
we use
\begin{equation} 
\Delta S_V = \int^T_{T_0} \frac{1}{T'} \left( \frac{\delta E}{\delta T'} \right)dT'
\end{equation}
where $E$ is the total internal energy which must be the sum of the
classical kinetic ($1.5Nk_BT$)  and  potential ($\langle U_{MX_2}\rangle$)
energy contributions for the $N$ particle system. Since the 
ensemble average of the potential energy for BKS silica as well
as TRIM BeF$_2$ can be fitted by functional form, $a(V)+ b(V)T^{3/5}$, the
integrand in equation (14) is simple to evaluate.

Since the internal energy estimates obtained using the short-range BKS and
the BKS potential with explicit Ewald summation used here are very small,
we used the same BLJ reference state as
used by Saika-Voivod and co-workers. The potential energy parameters
for this BLJ system, referred to as BLJ$_{SiO_2}$,  are listed in Table VI.
and the entropy of this system at 4000K and 2.307 g cm$^{-3}$ was
previously evaluated as 84.028 J mol$^{-1}$ K$^{-1}$. For this 
state point, we used equations (13) and (14) in conjunction with  molecular
dynamics simulations of systems at different values of $\lambda$ to compute the
entropy of BKS silica as 76.82 J K$^{-1}$, which differs by just 2.5\%
from the entropy of short-range BKS silica.  The entropies of BKS silica
over the entire range of temperature and density were then calculated
relative to this  state point.

Given the stoichiometry and the similar radius ratios of BeF$_2$ and SiO$_2$,
it was possible to  rescale the BLJ$_{SiO_2}$ in order to define an appropriate
reference system for BeF$_2$, denoted by BLJ$_{BeF_2}$. Since the locations of
the first minima  in the Be-Be, F-F and Be-F RDFs are
very similar to those in the corresponding RDFs for silica,
the length scale parameters for the BLJ$_{BeF_2}$ state system
were kept the same as for BLJ$_{SiO_2}$. The relative well-depths of the
three LJ-type pair interactions were also kept the same. 
Therefore if $\epsilon_{M-X}$ and $\sigma_{M-X}$ are chosen as the
units of energy and length respectively, then the two BLJ systems would
have identical internal energies and entropies at the same number density and
reduced temperature. To set a suitable energy scale for the BLJ system, we
use $\epsilon_{Be-F}\approx \epsilon_{Si-O} 
(T^{max}_{TMD,BeF_2})/(T^{max}_{TMD,SiO_2})$ and set 
$\epsilon_{Be-F}$ as 14.21K, as shown in Table VI. As discussed above, the
BLJ$_{SiO_2}$ liquid at 4000K and 2.307 g cm$^{-3}$ has total entropy
equal to 84.028 kJ mol$^{-1}$ K$^{-1}$= 10.106$k_B$, excess entropy of 9.23$k_B$ and internal
energy of 56.40$\epsilon_{Si-O}$. 
The equivalent state point for BeF$_2$ melt would be 
at 1777K and 1.805 g cm$^{-3}$ which would have the same internal
energy and excess entropy in reduced units. In order to calculate the
total entropy, the ideal gas contribution must be computed using masses
of Be and F. The total entropy of BLJ$_{BeF_2}$ was then found to be
70.7 J mol$^{-1}$K$^{-1}$=8.5$k_B$ at this state point.  The entropies
over the entire range of temperature and density  for BeF$_2$ were
then calculated relative to this  state point.

\section{Ionic Melts: Silica and Beryllium fluoride}

Figure 1 shows the thermodynamic excess entropy as a function of density
for SiO$_2$ and BeF$_2$ melts. Unlike the monotonic decrease in $S_e$
with $\rho$ seen in simple liquids, both systems have a well-defined 
anomalous regime where excess entropy rises with increasing density.
The anomalous behaviour is more pronounced for the low temperature
isotherms where the anisotropic interactions stabilizing local
tetrahedral order are large in comparison to thermal kinetic
energies.  All the $S_e(\rho )$ curves have a well-defined maximum
beyond which the excess entropy decreases with density; the maxima are
more pronounced for the low temperature isotherms with $T< T_{TMD}^{max}$
but can be identified in the high temperature isotherms as well. 

It is useful to compare the $S_e(\rho )$ behaviour with the
density dependence of the pair correlation entropy, $S_2$, for 
SiO$_2$\cite{scc06} and  BeF$_2$\cite{asc07}. The maxima in $S_e(\rho )$ and 
$S_2(\rho )$ curves  are located at essentially the same value of density
within simulation error. For example, maxima in $S_e$ coincides with the 
maxima observed in $S_2$ at $\rho = 3.4$g cm$^{-3}$ at 4000 K for SiO$_2$ and 
at $\rho = 2.82$ g cm$^{-3}$ at 1500 K for Be$F_2$. The most notable
qualitative difference between the $S_e(\rho )$ and $S_2(\rho )$ curves
is the absence of a minimum in the $S_e$ curves at low densities.
The $S_2 (\rho )$ curves for all waterlike systems show a minimum
at low densities which coincides with the maximum in the tetrahedral
order parameter along the isotherms, indicating that $S_2$ is
more sensitive to tetrahedral order than $S_e$.

Figure 2 shows the temperature dependence of the excess entropy of
BeF$_2$ and SiO$_2$ along different isochores. We plot $S_e$ as a 
function of $T^{-2/5}$ since earlier density functional as well
as simulation studies suggest that the excess entropy of simple liquids,
even within the pair correlation approximation, obeys a $T^{-2/5}$ scaling
\cite{rt98,cc07}.
Figures 2(a) and (b) show that both $SiO_2$ and $BeF_2$ melts show an
approximate $T^{(-2/5)}$ scaling, with small deviations from the
scaling behaviour in the anomalous regime.  Note that a $T^{3/5}$ scaling
predicted for the configurational potential energy is obeyed by both the
melts and has been used when performing thermodynamic integration 
\cite{pha97}.
Interestingly, the pair correlation entropy, $S_2$, plotted as a function of
$T^{-2/5}$, of liquids with waterlike
anomalies  shows a significant change between on going from the normal to 
the structurally anomalous regime.

To summarise, the qualitative differences that emerge in the density and
temperature dependence of $S_e$ and $S_2$ are: (i) absence of a minimum in
$S_e$, unlike the minimum in $S_2$ which correlates with the maximum in
tetrahedral order and (ii) the temperature dependence of $S_e$ along isochores
is well described by $T^{-2/5}$ scaling across both the normal and anomalous
regimes, unlike in the case of $S_2$. This suggests that it is
important to examine the relationship between the residual multiparticle
entropy, $\Delta S$, and tetrahedral order parameter, $q_{tet}$, as is
done below.

Figure 3(a)  displays the correlation between pair correlation entropy
$S_2$ and thermodynamic excess entropy, $S_e$, in the case of of SiO$_2$.
A strong correlation  between $S_2$ and $S_e$ is seen for the low isotherms
of T = 5000, 4500 and 4000 K i.e. for $T/T_{TMD}^{max} \leq 1$. 
For temperatures greater than $T_{TMD}^{max}$ (T = 5500 and 6000 K), 
the $S_2$ versus $S_e$ curve shows two branches, a high density and a 
low density one. The density demarcating the two branches corresponds to
3.0 g cm$^{-3}$.  Figure 3(b) shows the $S_2$ versus $S_e$
correlation plot for BeF$_2$. Isotherms above 2000K ($T_{TMD}^{max}=2310K$),
the curves have distinct high and low density branches with the demarcation
density being 2.4 g cm$^{-3}$.

In order to understand the relationship between $S_e$ and $S_2$, we 
plot $\Delta S$ as a function of $\rho$ in Figure 4.
Several interesting features of the RMPE as a function
of density are common to both BeF$_2$ and SiO$_2$ melts. In both
systems, $\Delta S$ is essentially constant after a characteristic
density $\rho_{TMD}^{max}$ which marks the maximum density for which
the thermodynamically anomalous behaviour in the density is observed
in the liquid phase.
This corresponds to a density of 3 g cm$^{-3}$ for SiO$_2$ and 2.4
g cm$^{-3}$ for BeF$_2$. In both systems isotherms which lie
at or below $T_{TMD}^{max}$ show a decrease in $\Delta S$ with increasing
$\rho$ till $\rho_{TMD}^{max}$ is reached. In contrast, isotherms which
lie above $T_{TMD}^{max}$ show an increase in $\Delta S$ till 
$\rho_{TMD}^{max}$ is reached.

Figure 4 also shows that the RMPE is negative for all state points in the
case of SiO$_2$ but not in the case of BeF$_2$. This may in part be due
to the fact that the BKS interaction potential in the case of SiO$_2$ 
does not include any explicit cation-cation (or Si-Si) interactions.
We find that the RMPE contributes about 10\% or less to the excess
entropy, $S_e$, of the ionic liquids in the normal regime. In the anomalous
regime, specially at low temperatures, the contribution of the $\Delta S$
terms can be as large as 30\% of the total excess entropy in terms of
the absolute magnitudes.

To assess the quantitative reliability of the pair correlation approximation
to obtain the total thermodynamic entropy, Figure 5 shows the
quantity $\vert\Delta S\vert /S\%$ where $S=S_{id}+ S_e$ is the total 
thermodynamic entropy. The error caused  by the pair correlation approximation
is of the order of 6\% or less for both ionic melts in the normal regime,
but as expected from the behaviour of the RMPE, the error can be
greater in the anomalous regime. Other than the 1500K isotherm of
BeF$_2$, the magnitude of the error for all state points studied 
is less than 10\%.

The above results  suggest that
the nature of the relationship between the residual multiparticle
entropy and structural order is significantly different in the normal
and anomalous regime. Therefore, we plot $\Delta S$  as a function
of the tetrahedral order parameter $q_{tet}$ in Figure 6. The 
local tetrahedral order parameter, $q_{tet}$, associated
with  an atom $i$   is  defined as
\be
q_{tet} = 1 -\frac{3}{8}\sum_{j=1}^3\sum_{k=j+1}^4 (\cos \psi_{jk}+ 1/3)^2
\ee
where $\psi_{jk}$ is the angle between the bond vectors {\bf r}$_{ij}$
and {\bf r}$_{ik}$ where $j$ and $k$ label the four nearest neighbour
atoms of the same type \cite{ed01}.  
Figure 6 shows that $\Delta S$ has a strong
negative correlation with tetrahedral order in this regime. Clearly, the
dominant contribution to the RMPE in the anomalous regime is from the
three-body terms which are  strongly anti-correlated with tetrahedral order.

We now reconsider the behaviour of $S_2$ versus $S_e$ at temperatures
less than or equal to $T_{TMD}^{max}$ in Figure 3. $S_2$ and $S_e$
are strongly correlated and increase with increasing density till
$\rho_{max}$ is reached. 
The anomalous regime also shows a strong correlation between  tetrahedral
order and $S_2$ or any related measure of structure in the pair correlation 
function \cite{ed01}.
The effect of the local tetrahedral ordering, imposed in the case of
ionic melts from the steric factors associated with the relative ionic
radii, is to strongly couple the two and three-body contributions to
the excess entropy.
For isotherms lying above $T_{TMD}^{max}$, at low densities $S_2$ is
essentially constant while $S_e$ increases. The multiparticle
correlations in this case serve to increase the entropy with
increasing density and are essentially uncorrelated with the
tetrahedral order. For densities above $\rho_{max}$ and temperatures
above $T_{TMD}^{max}$,
the system behaves essentially as a simple liquid, with $S_2$ and
$S_e$ both showing a negative correlation with density though with
a relatively small variation in either entropy or local order.

\section{Model Potentials for Water}

We first consider the the pair correlation estimate of
the entropy, based on the $g_{OO}(r)$, $g_{OH}(r)$ and $g_{HH}$ pair 
correlation functions, with the total entropy as estimated from
thermodynamic integration for the SPC/E model for water.
The experimentally obtained $g_{OH}(r)$ and $g_{HH}(r)$ RDFs will show a 
sharp, narrow peak corresponding to the intramolecular distances between
these atoms. In the case of rigid body molecular dynamics, this peak will
reduce to a $\delta$-function. The $S_2$ contribution can be computed from
equation (3) from the $g(r)$ functions without 
this additional $\delta$-function contribution.

Figure 7(a) shows the density dependence of  the thermodynamic entropy, $S^*$,
taken from ref.\cite{sslss00}, for several isotherms of SPC/E
water lying between 210K and 300K.
Comparison with the pair correlation estimate,
$S^*\approx S_{id} + S_2$, shown in Figure 7(b), 
indicates that the pair correlation entropy shows the correct qualitative
behaviour. 
The anomalous rise in entropy  between 0.95 and 1.10 
g cm$^{-3}$ is seen in both Figures 7(a) and 7(b) for isotherms at
210,220 and 230 K. The quantitative errors introduced by the
pair correlation approximation for SPC/E water are assessed by plotting
$\vert \Delta S\vert /S \%$ where $S$ is the thermodynamic excess entropy taken
from ref.\cite{sslss00} as a function of density in Figure 7(c).
The errors due to the pair correlation approximation at temperatures
of 240K and above lie between 20 and 30\% and are almost independent
of density. Not surprisingly, this is larger than errors observed
for the ionic melts in the normal liquid regime. At low temperatures,
specially in the anomalous regime, there is a strong density dependence
of the errors; for example, for the 210K isotherm, the error varies 
between 10 and 60\% .

Since the pair correlation entropy provides a very reasonable
estimate of the temperature-density regime in which thermodynamic
anomalies may  be expected, we consider the two other effective pair
potential models of water known to have very different regimes of
anomalous behaviour (Table I).  The TIP5P model reproduces the experimental
data closely in this respect and shows a $T_{TMD}^{max}$ value of 282K.
In the case of the SPC/E model, however, the  anomalous
regime is shifted to temperatures approximately 30 to 40K below those
see experimentally.  In the case of TIP3P, the anomalies appear to occur
at even lower  temperatures though a detailed mapping of waterlike anomalies
has not been performed for this model. The TIP3P model is, however, of
considerable interest since it is frequently used to model the aqueous
solvent in biomolecular simulations \cite{jcmik83}. To illustrate
the quantitative differences between the models, one may note that
a recent study indicates that the TMD at 1 atm pressure occurs at 241K,
182K and 285K for the SPC/E, TIP3P and TIP5P models respectively \cite{va05}.
Figure 8 shows the density dependence of the  pair corrrelation entropy, 
$S_2$, as a function of density for different isotherms for all three models.
It is immediately apparent that the thermodynamically anomalous
regime can be correctly identified for all three models from
this pair correlation information. An excess entropy anomaly exists in
the TIP3P, SPC/E and TIP5P models below 200K, 260K and 360K 
respectively. The strength of the excess entropy
anomaly, as identified by $(\partial S_2 /\partial (\ln\rho ))_T$ \cite{etm06}, 
is maximal in the TIP5P model and least in the TIP3P model.

\section{Conclusions}

This paper compares the pair correlation entropy ($S_2$), determined from the 
atom-atom radial distribution functions  with the excess entropy of two
ionic melts (liquid silica and beryllium fluoride) and a molecular liquid 
(water).  The three liquids that we have chosen to study show distinct normal 
and anomalous regimes. The anomalous regime is characterised by significant
departures from simple liquid behaviour and shows a rise in thermodynamic
entropy on isothermal compression. This anomalous entropy behaviour can
be connected to the existence of waterlike structural, kinetic and
thermodynamic entropies. The pair correlation entropy is sufficiently
accurate that the density-temperature regime of anomalous behaviour
as well as the strength of the entropy anomaly 
can be predicted reliably for both ionic melts as well as different
rigid-body pair potentials for water. 
This simple connection between atom-atom radial distribution functions
and thermodynamic anomalies predicted by different water models has not
been discussed previously in the literature.

To assess the quantitative contribution of
the pair correlations to the thermodynamic entropy, we compare the 
RDF-based $S_2$ estimator for ionic melts with the thermodynamic excess entropy,
$S_e$, measured with respect to an ideal gas mixture. Errors in the total
thermodynamic entropy for ionic melts  due to the pair
correlation approximation are of the order of 6\% in the normal
regime but can be significantly larger in the anomalous regime.
In the case of water, we compared the total thermodynamic entropy with the
estimate based on the atom-atom RDFs. As expected given the rigid-body
constraints for molecular liquids, the pair correlation approximation
then causes significantly larger errors, between 20 and 30\%, in the
normal liquid regime.

In the case of ionic melts, the high density limit of the
anomalous regime is marked by the maximum in the $S_e(\rho )$ curves
at a given temperature and this is well reproduced by the 
$S_2 (\rho )$ curves. The excess entropy curve, $S_e (\rho)$, does
not show a minimum,  unlike the $S_2(\rho )$ curves for which the minimum
correlates with the maximum in
tetrahedral order. Along an isochore, the temperature dependence of $S_e$ is
well described by $T^{-2/5}$ scaling across both the normal and anomalous
regimes.  In contrast, $S_2$ shows $T^{-2/5}$ scaling behaviour only in
the normal liquid regime.
	
Comparison of pair correlation approximation to the excess entropy
from scattering data with calorimetric estimates of the entropy can
provide interesting insights into the role of multiparticle
interactions in different liquids. For example, 
in the case of ionic melts, we show that the relationship between $S_2$ and 
$S_e$ is significantly different in the normal and anomalous regime. 
Strong, local anisotropic interactions are dominant in the anomalous regime
and serve to couple the two- and three-body contributions to the entropy.
Thus the correlation between local order metrics and RMPE in 
simple and anomalous liquids is qualitatively different.
	
{It is useful at this point to compare our results for 
excess entropy of the three tetrahedral
network-forming liquids with the earlier results for H$_2$O using an ideal gas 
of rigid rotors as a reference state \cite{lk96,ssg03,essg06}. Clearly the 
separation of time scales between intramolecular and intermolecular 
interactions of water justifies the rigid-body ideal gas limit as the 
physically more reasonable choice. In the case of ionic melts, the molecular 
limit is not an obvious choice and defining the reference state as a 
non-interacting mixture of atoms is a reasonable option. The greater accuracy 
of the pair correlation estimator for the excess entropy of ionic melts,
as compared to water, is a consequence of the appropriateness of
choice of reference state. Both approaches indicate that the
contribution of multiparticle interactions to the excess entropy changes
qualitatively on going from the normal to the anomalous regime.
The translational entropy, as defined in refs.\cite{lk96,essg06},
corresponds to the contribution of the O-O pair correlation function
to the pair correlation entropy as defined in our work. The orientational
entropy must  correspond to a subset of  the three-body terms
in our  formulation which is supported by the strong correlation of the
RMPE with the tetrahedral order parameter.}

In conclusion, the results presented in this paper suggest that atom-atom 
radial distribution functions can be used to construct a reliable, 
semiquantitative structural
estimator for the entropy. The comparison of such a structural estimator
for the entropy with calorimetric data could lead to interesting insights
into the role of pair and higher-order particle correlations in determining
thermodynamic and transport properties  of liquids  and serve as 
additional inputs for improving structure prediction by reverse 
Monte Carlo or related techniques based on RDF data.  We also note that
there has been considerable interest in developing coarse-graining
strategies based on effective pair potentials  for complex liquids.
For example, isotropic models of water which reproduce the $g_{OO}(r)$
function  have been constructed but do have some problems with
representability and transferability \cite{aal02,jhl07,hs93,lh06,mrytd07}. 
It may be interesting to consider
coarse graining strategies which modify the RDF data to reproduce the
excess entropy of the liquid.

{\bf Acknowledgements} This work was financially 
supported by the Department of Science and Technology, New Delhi.
Computational support from the Computer Services Centre are gratefully
acknowledged.
MA and RS thank Indian Institute of Technology Delhi and 
Council for Scientific and Industrial Research
respectively for financial support.
The authors would like to thank Ivan
Saika-Voivod for clarifications with regard to their published work.

\newpage
\begin{table}
	\caption{Maximum temperature and corresponding density for locus
	of state points corresponding to temperatures of maximum density
	(TMD) for different isobars. These temperatures and densities
	mark the maximum in the density-temperature ($\rho-T$)-plane
	of the curve connecting state points with 
	$(\partial V/\partial T)_P=0$. The data
	for $T_{TMD}^{max}$ and $\rho_{TMD}^{max}$ were taken from
	the literature and the apprpriate reference is cited in the
	column heading.}

	\begin{tabular}{ccccc}
		\hline
		\qquad & SPC/E\cite{ed01} & TIP5P\cite{sbgn03,ymss02} & BeF$_2$\cite{asc07} & SiO$_2$\cite{sdp02} \\
		\hline
		$T_{TMD}^{max}$/K & 249 &  282 & 2310 & 4940 \\
		$\rho _{TMD}^{max}$/ g cm$^{-3}$ & 0.97 & 1.0 & 1.805 & 2.307\\
		\hline
	\end{tabular}
\end{table}

\newpage

\begin{table}
	\caption {Potential parameters for BKS SiO$_2$ with 30-6 Lennard-Jones 
	correction terms \cite{bks90,kfb91,ssp00}.}
\begin{tabular}{cccccc}
\smallskip\\
\hline\hline
~&~&~&~&~\\
$i-j$  & $A_{ij}$  & $b_{ij}$  & $C_{ij}$  & $\epsilon_{ij}$  &
 $\sigma_{ij}$ \\
\   &  (kJ mol$^{-1}$) &  ($\mathring{A}^{-1}$) &  (kJ mol$^{-1}$) &  (kJ mol$^{-1}$) &  ($\mathring{A}$)\\
~&~&~&~&~\\
\hline
~&~&~&~&~\\
$O-O$  & 134015   & 2.76  & 16887.3 & 0.101425  & 1.7792 \\
$Si-O$ & 1737340  & 4.87  & 12886.3 & 0.298949  & 1.3136 \\
~&~&~&~&~\\
\hline\hline
\end{tabular}
\end{table}

\newpage
\begin{table}
\caption{TRIM potential parameters for \bef2 \cite{wac76,hma01,ha97}}
\label{paramstable}
\begin{tabular}{cccccccc}
\hline\hline
~$\sigma^+$ & ~$\sigma^-$ & $\rho$ & $z^+$ & $z^-$ & $n^+$ & $n^-$ & $b$ \\
($\mathring{A}$) & ($\mathring{A}$) & ($\mathring{A}^{-1}$) & ~ & ~ & ~ & ~ & (kJ mol$^{-1}$) \\
\hline
~&~&~&~\\

0.93 & 1.24 & 0.29 & 2 & -1 & 2 & 8 & 34.33\\
~&~&~&~\\
\hline\hline

\end{tabular}
\end{table}

\newpage
\begin{center}
\begin{table}
\caption
{Comparison of parameters of common water models
\cite{jcmik83,bgs87,mj00}.  SPC/E and TIP3P are three site models 
with the respective charges and masses centered at the O and H sites 
\cite{bgs87,jcmik83}. 
TIP5P is a five site model with  two additional massless sites M \cite{mj00}.}
\setlength{\tabcolsep}{6mm}

\begin{tabular}{lllllll}   \hline
                         &      SPC/E       &      TIP3P    & TIP5P    \\
\hline
$\epsilon$ (kcal mol$^{-1}$)    &   0.155   &      0.152    &   0.160  \\
$\sigma(\mathring{A})$    &         3.166   &      3.150    &   3.120  \\
$r_{OH}$                  &         1.000   &      0.9572   &   0.9572 \\
$\angle{HOH(deg)}$        &      109.47     &      104.52   &  104.52  \\
$q_{O}$                   &     -0.8472     &     -0.834    &   0.0      \\
$q_{H}$                   &      0.4238     &      0.417    &   0.241   \\
$q_{M}$                   &      -          &      -        &  -0.241   \\
$r_{OM}(\mathring{A})$    &      -          &      -        &   0.70     \\
$\angle{MOM(deg)}$        &      -          &      -        &  109.47    \\
\hline
\end{tabular}
\end{table}
\end{center}

\newpage
\begin{center}
	\begin{table}
\caption{Computational details of molecular dynamics simulations of
SiO$_2$, BeF$_2$ and H$_2$O. The simulation cell size is given
in terms of the number of formula units, $M$, present in the system.
An MD time step of 1 fs was used for all three systems. The time
constant for the Berendsen thermostat is denoted by $\tau_B$.
Equilibriation and production run lengths are denoted by $t_{eq}$ and
$t_{prod}$ respectively.}
\bigskip
\begin{tabular}{ccccccl}
\smallskip\\
\hline\hline
System & Model & M &   $t_{eq}$ & $t_{prod}$ & $\tau_B$ & MD algorithm\\ 
 &  &  & (ns) & (ns) & (ps) & \ \\
\hline
SiO$_2$ & BKS  & 150 &  3-6 & 5-10 & 200 & Verlet\\
BeF$_2$ & TRIM & 150 &  4-8 & 6-8 & 200 & Verlet\\
H$_2$O & SPC/E & 256 &  1 & 1 & 10 & Verlet + Quaternion\\
H$_2$O & TIP3P & 256 &  0.25-0.5 & 0.5-0.75   & 1& Verlet + SHAKE\\
H$_2$O & TIP5P & 256 &  0.25-0.5 & 0.5-0.75   & 1 & Verlet + Quaternion\\
\hline
\end{tabular}
\end{table}
\end{center}
\newpage

\begin{table}
	\caption{Potential energy parameters for the binary Lennard-Jones
	fluid used as reference state for thermodynamic integration to obtain
	entropies of BKS SiO$_2$ and TRIM BeF$_2$.}
	\begin{tabular}{ccccccc}
	\hline
	\multicolumn{3}{c}{SiO$_2$} & \quad & \multicolumn{3}{c}{BeF$_2$}\\
	\hline
	$i-j$  & $\epsilon$ & $\sigma$d & \quad & $i-j$  & $\epsilon$ & $\sigma$ \\
	\quad  & (kJ mol$^{-1}$) & ($\mathring{A})$ &\quad & \quad  & (kJ mol$^{-1}$) & ($\mathring{A}$)\\ 
	\hline
	Si-O  & 32.0  & 1.6 & \quad & Be-F & 14.21 & 1.6 \\
	Si-Si & 23.0  & 3.3 & \quad & Be-Be & 10.22 & 3.3\\
	O-O   & 23.0  & 2.8 & \quad & F-F   & 10.22 & 2.8\\
	\hline
\end{tabular}
\end{table}

\begin{center}
{\bf Figure Captions}
\end{center}

\begin{enumerate}

\item Thermodynamic excess entropy, $S_e$, as a function of density for
	(a) SiO$_2$ and (b) BeF$ _2$. The vertical line shows the position of 
	minima in the  $S_2(\rho )$ curves for SiO$_2$ \cite{scc06} and
	BeF$_2$ \cite{asc07}. 
	Unless otherwise stated, entropy is reported in units of $k_B$ per 
	atom in all the figures.  
	Isotherms are labelled in degrees K in all the figures.

\item Temperature dependence of the thermodynamic excess entropy, $S_e$, for 
 (a) SiO$_2$ and (b) BeF$_2$ along different isochores. The isochores
 are labelled by the density in g cm$^{-3}$.

\item Correlation between the pair correlation entopy, $S_2$, 
and the thermodynamic excess entropy, $S_e$, for (a) SiO$_2$ and (b) BeF$_2$. 
The vertical arrows indicate the lowest density 
 for each isotherm studied in this work.

\item Variations in residual multiparticle entropy, $\Delta S$, with density, 
$\rho$, along different isotherms, labelled by temperature in Kelvin, for 
(a) $SiO_2$ and (b) $BeF_2$.
	$\Delta S$ is reported
	in units of $k_B$ per atom. 

\item Percentage contribution of residual multiparticle entropy to
	the total thermodynamic entropy, $\vert\Delta S\vert/S$\%, along
	different isotherms for (a) $SiO_2$ and (b) $BeF_2$.

\item Correlation between the tetrahedral order parameter, $q$ and the 
residual multiparticle entropy, $\Delta S$, for (a) $SiO_2$ and (b) $BeF_2$.
	$\Delta S$ is reported
	in units of $k_B$ atom. 
	The arrows indicate the lowest density state point along an
	isotherm simulated in this study.

\item Total entropy of SPC/E water as a function of density. (a) Total 
thermodynamic entropy, $S$ taken from ref. \cite{sslss00} and (b) total 
pair correlation entropy, $S^* = S_{Id} + S_2$ calculated by considering water 
as a binary mixture of H and O atoms. From top to bottom, the isotherms 
correspond to T = 300($\bigtriangledown$), 260($\blacktriangle$), 
240($\bigtriangleup$), 230($\bullet$), 220($\bigcirc$) 
and 210 K($\blacksquare$). 
Entropies are reported in units of $k_B$ per 
atom.

\item Pair correlation entropy as a function of density for different
	model potentials of water: (a) SPC/E (b) TIP3P and (c) TIP5P.
Entropies are reported in units of $k_B$ per atom.

\end{enumerate}

\newpage
\clearpage

\begin{figure}
\caption{\quad}
\centering
\includegraphics[scale=0.75]{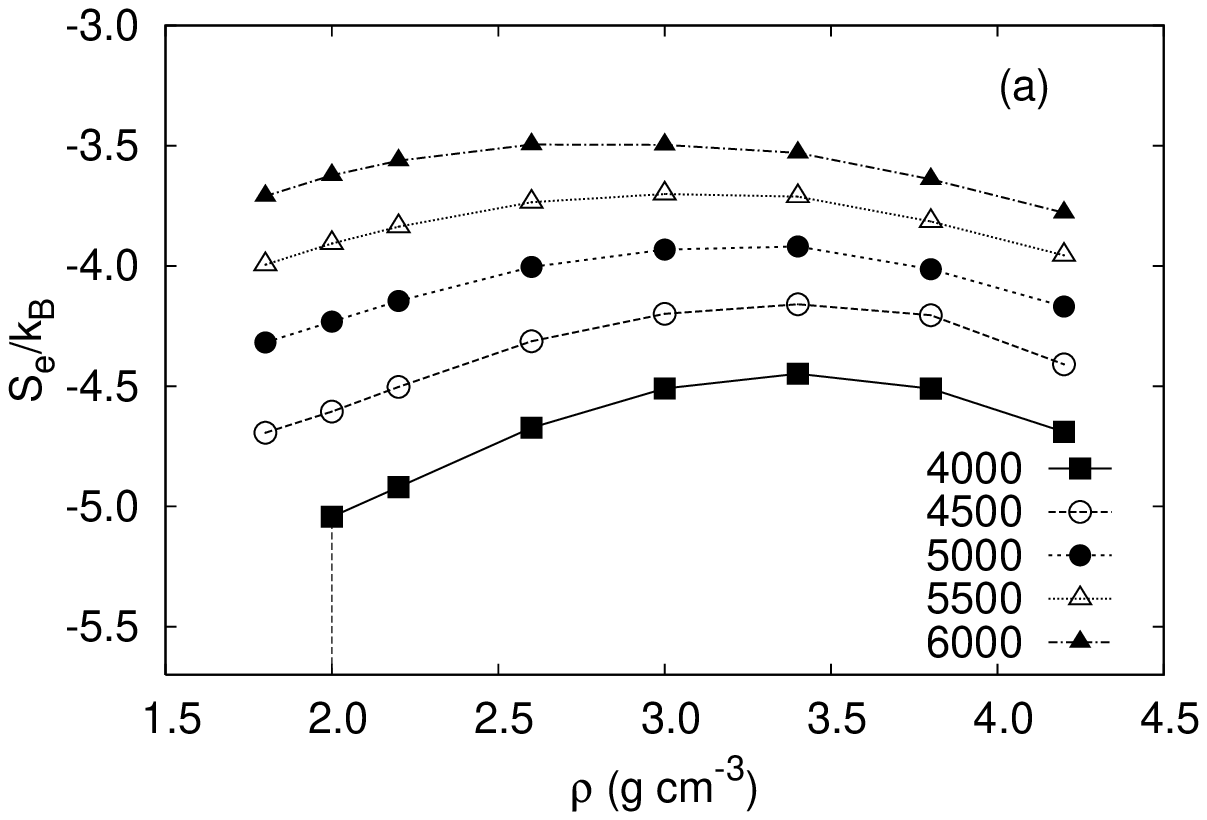}
\end{figure} 

\begin{figure}
\centering
\includegraphics[scale=0.75]{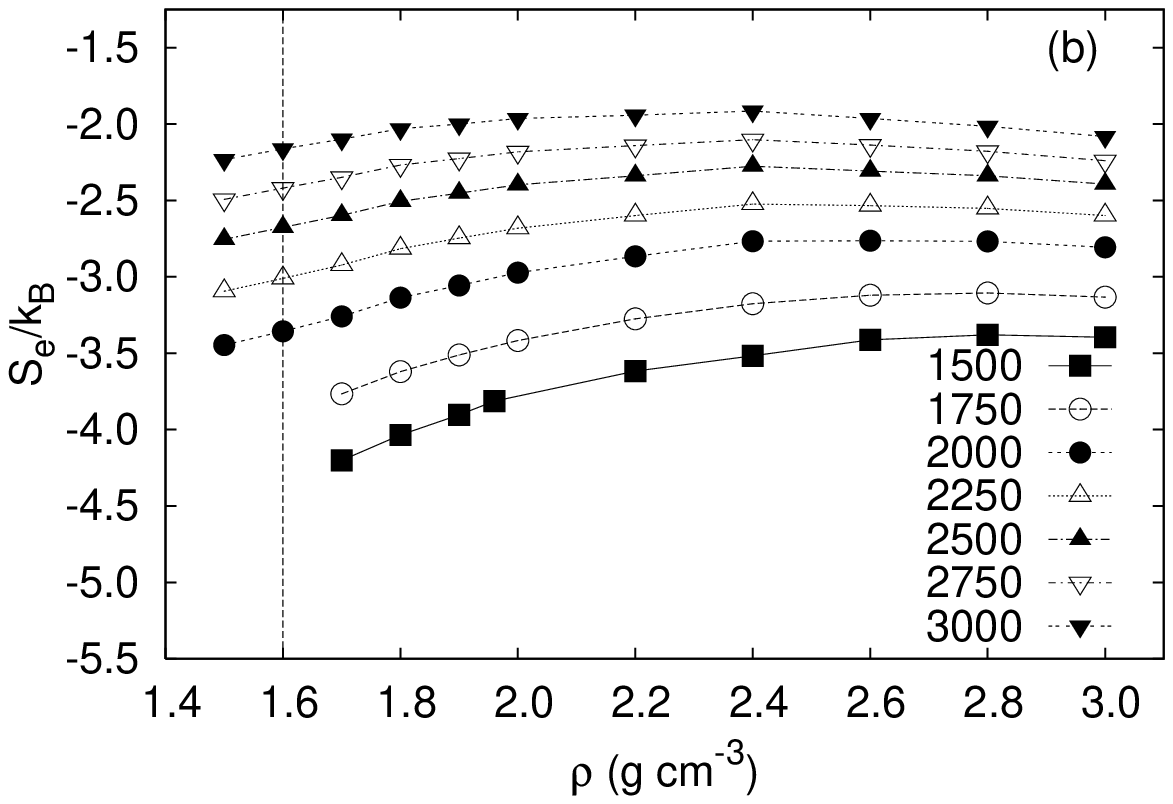}
\end{figure}

\newpage
\clearpage

\begin{figure}
\caption{\quad}
\centering
\includegraphics[scale=0.75]{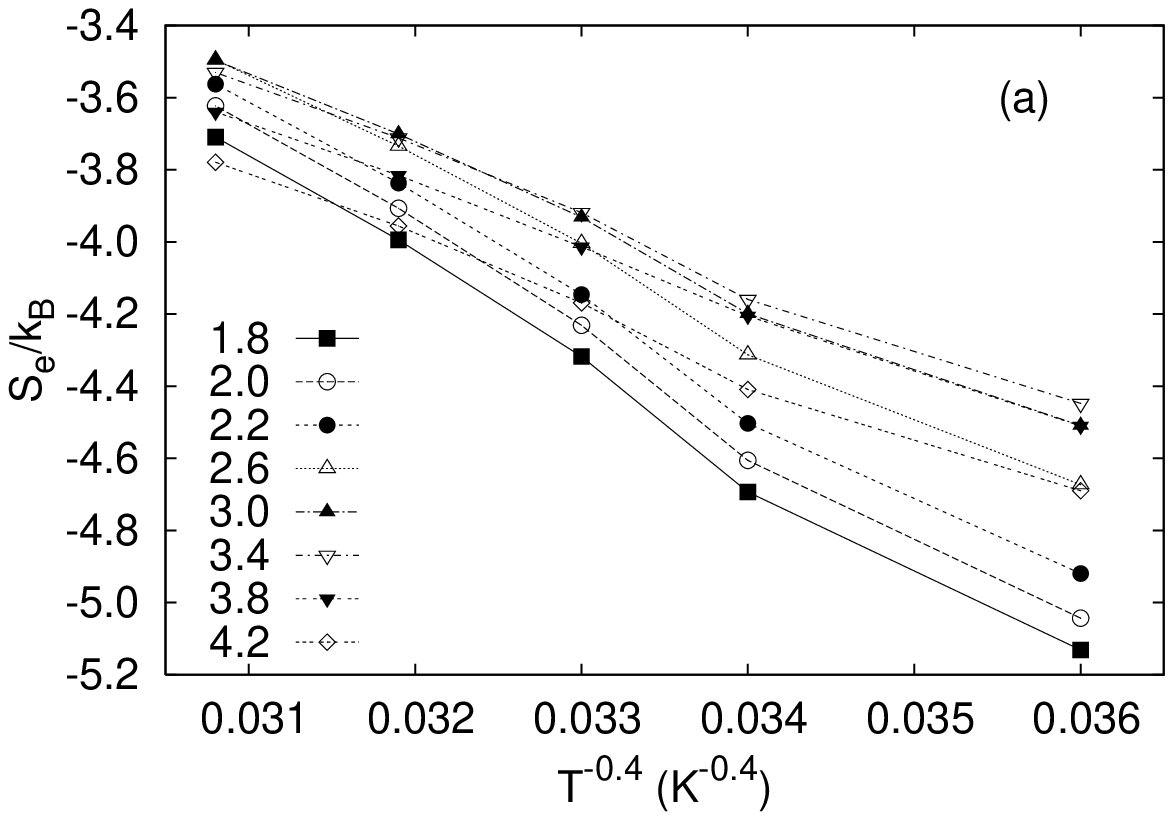}
\end{figure}

\begin{figure}
\centering
\includegraphics[scale=0.75]{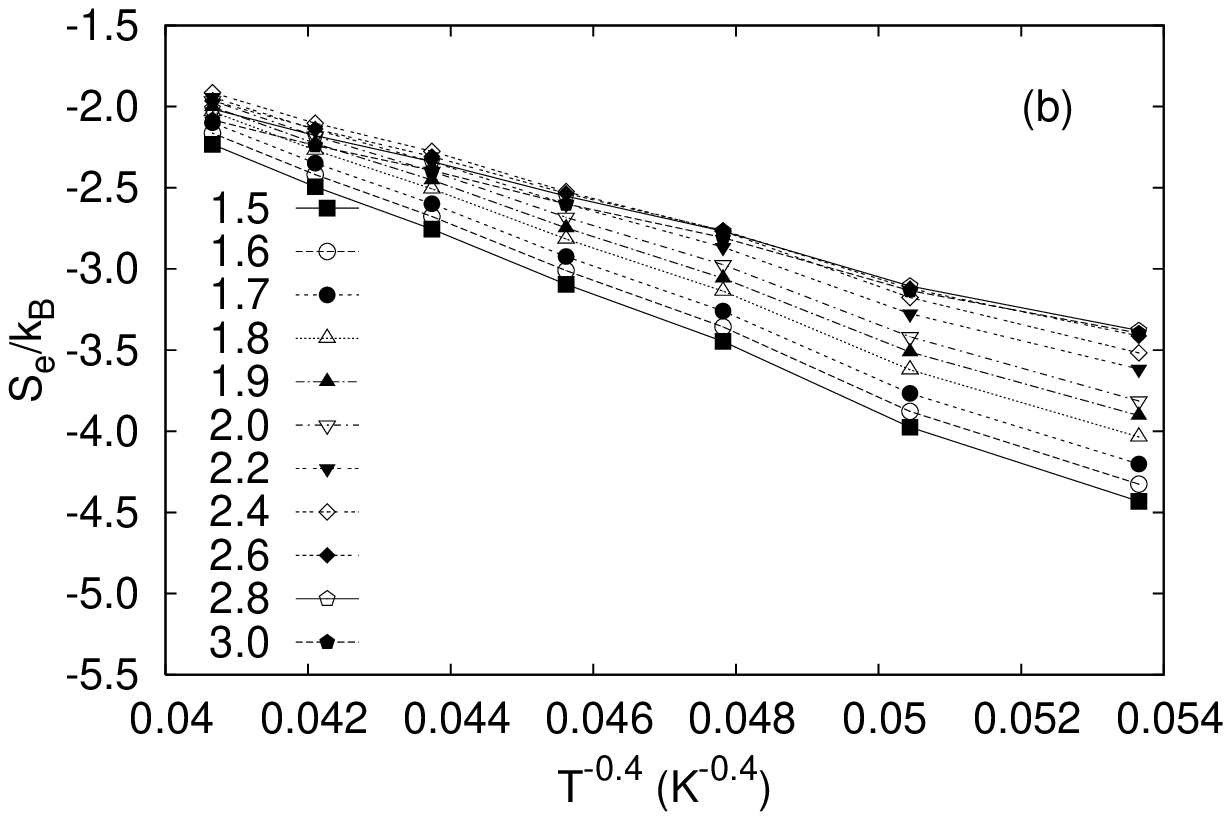}
\end{figure}

\newpage
\clearpage

\begin{figure}
\caption{\quad}
\centering
\includegraphics[scale=0.75]{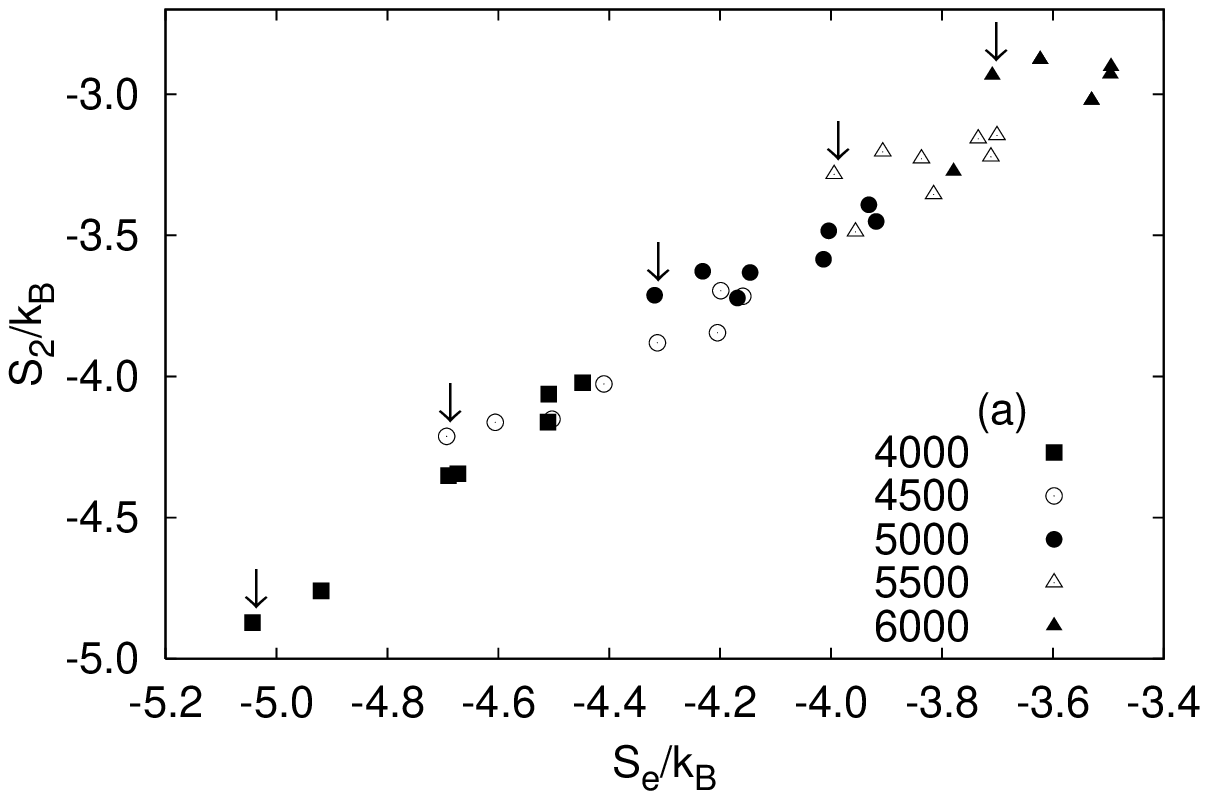}
\end{figure}

\begin{figure}
\centering
\includegraphics[scale=0.75]{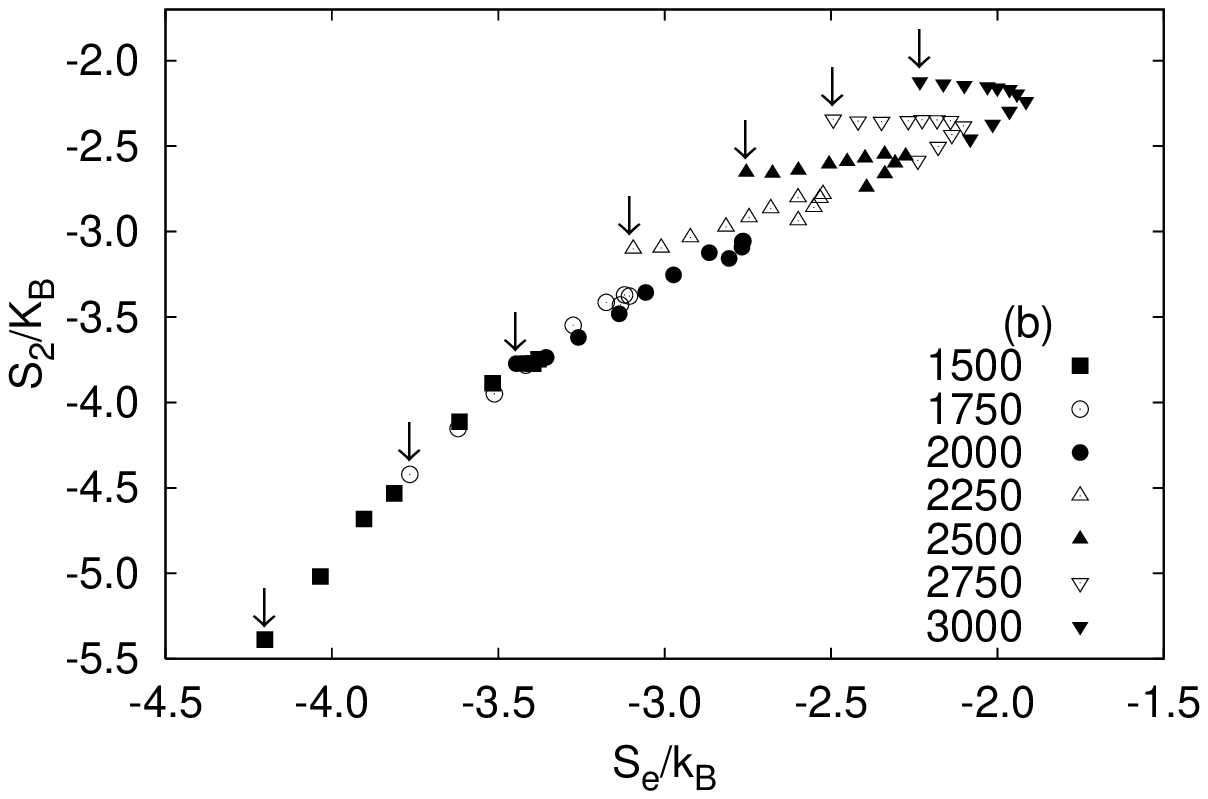}
\end{figure}

\newpage
\clearpage

\begin{figure}
\caption{\quad}
\centering
\includegraphics[scale=0.75]{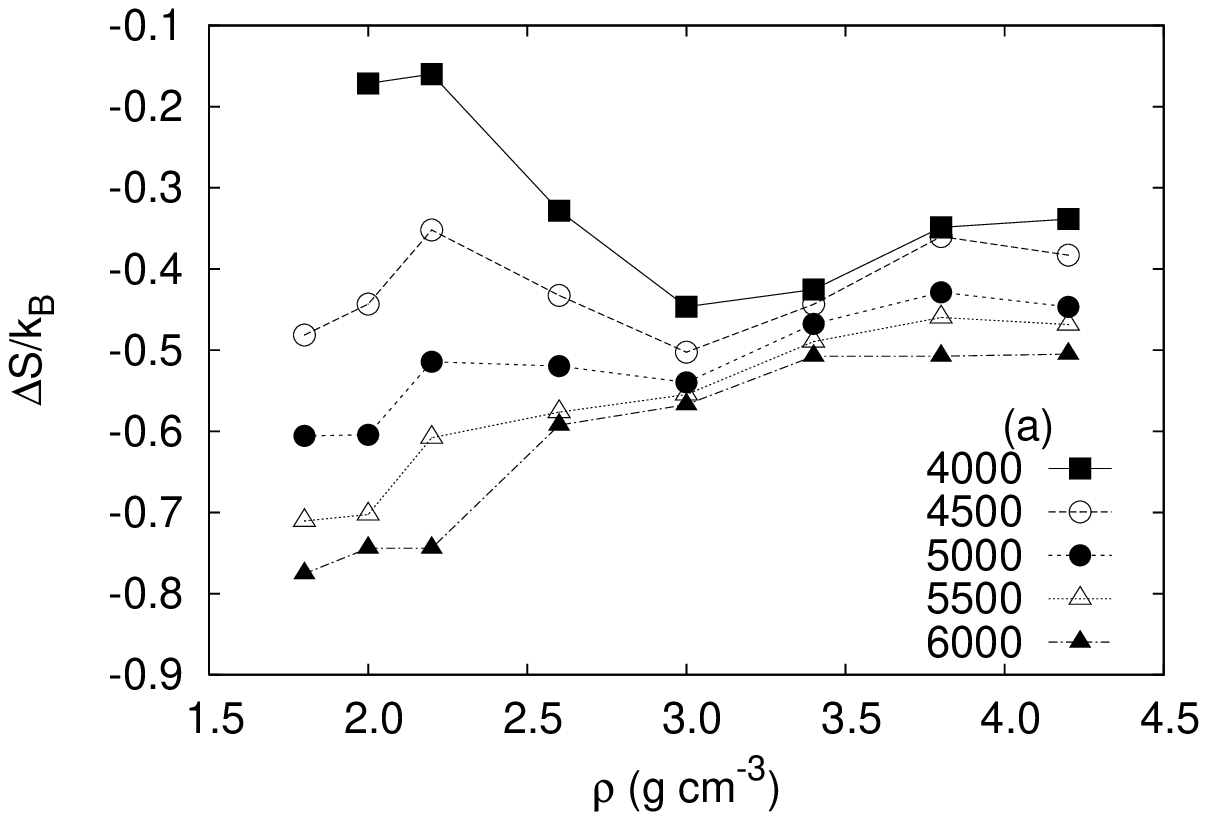}
\end{figure}

\begin{figure}
\centering
\includegraphics[scale=0.75]{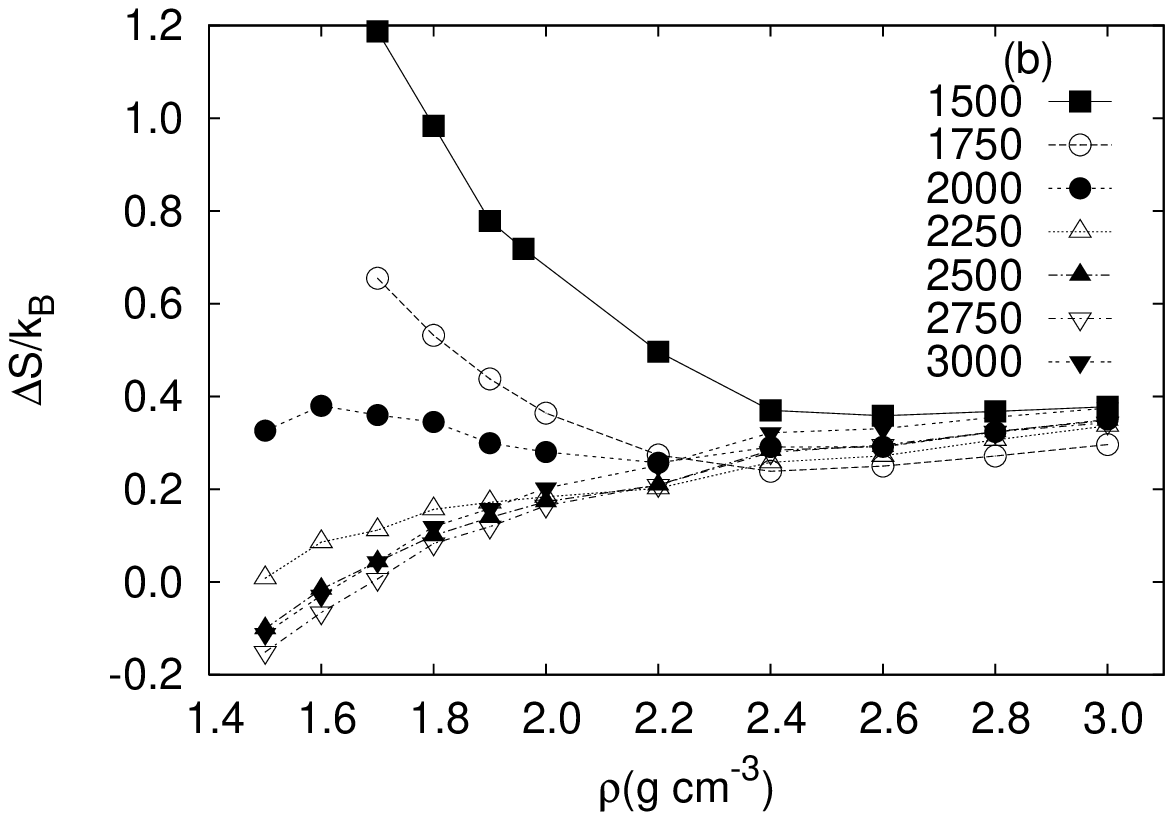}
\end{figure}

\newpage
\clearpage

\begin{figure}
\caption{\quad}
\centering
\includegraphics[scale=0.75]{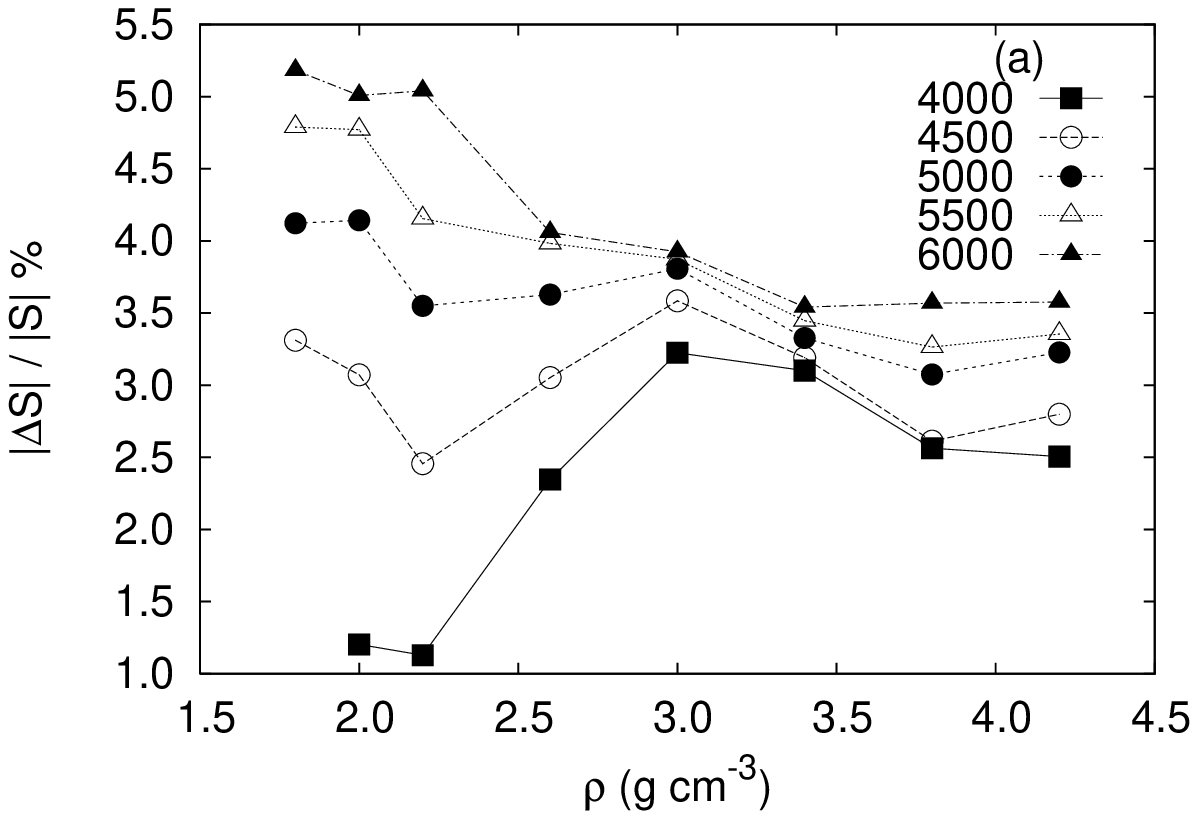}
\end{figure}

\begin{figure}
\centering
\includegraphics[scale=0.75]{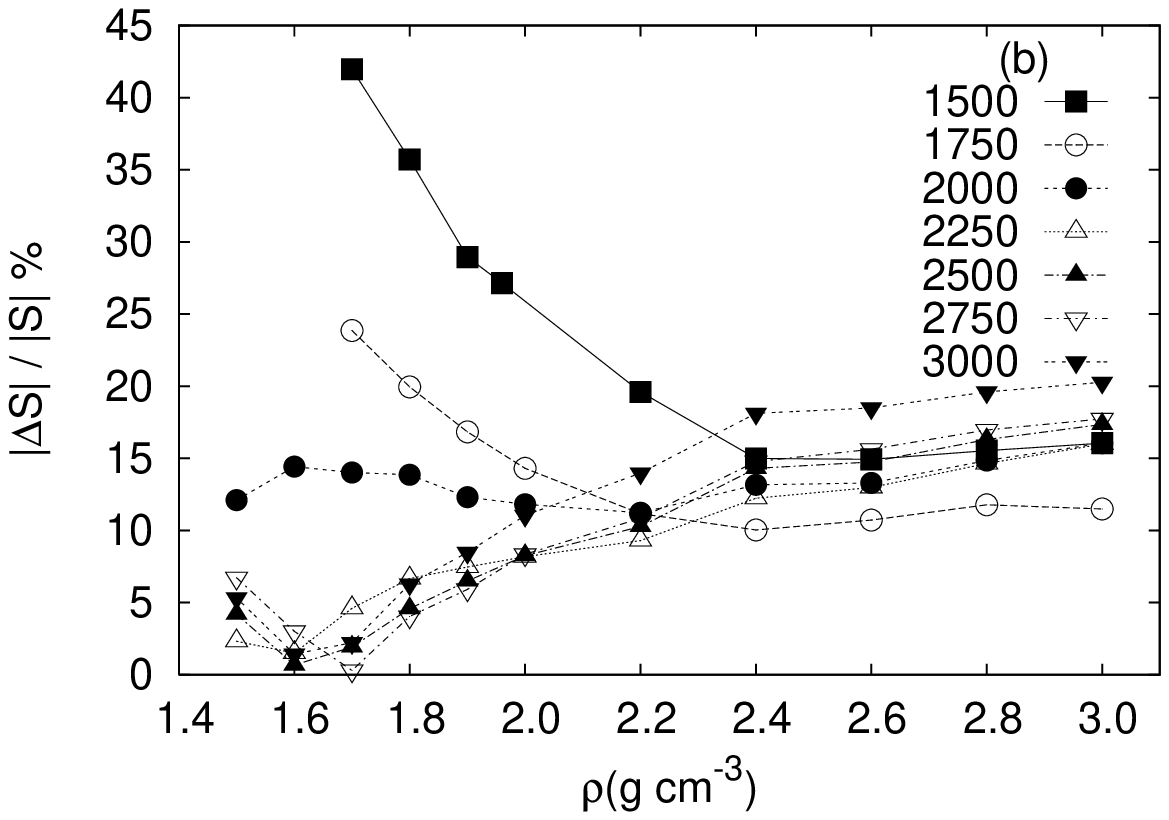}
\end{figure}

\newpage
\clearpage

\begin{figure} \caption{\quad} 
\includegraphics[scale=0.75]{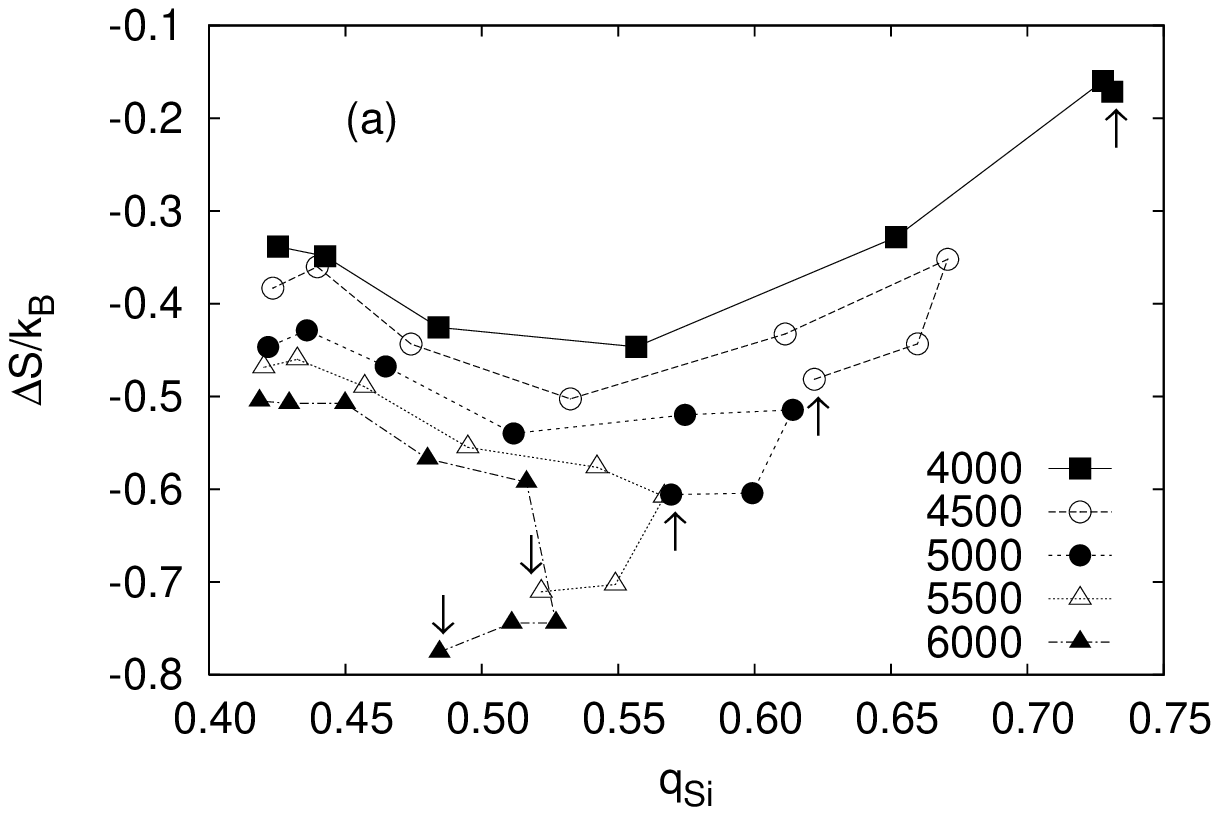}
\end{figure}

\begin{figure}
\centering
\includegraphics[scale=0.75]{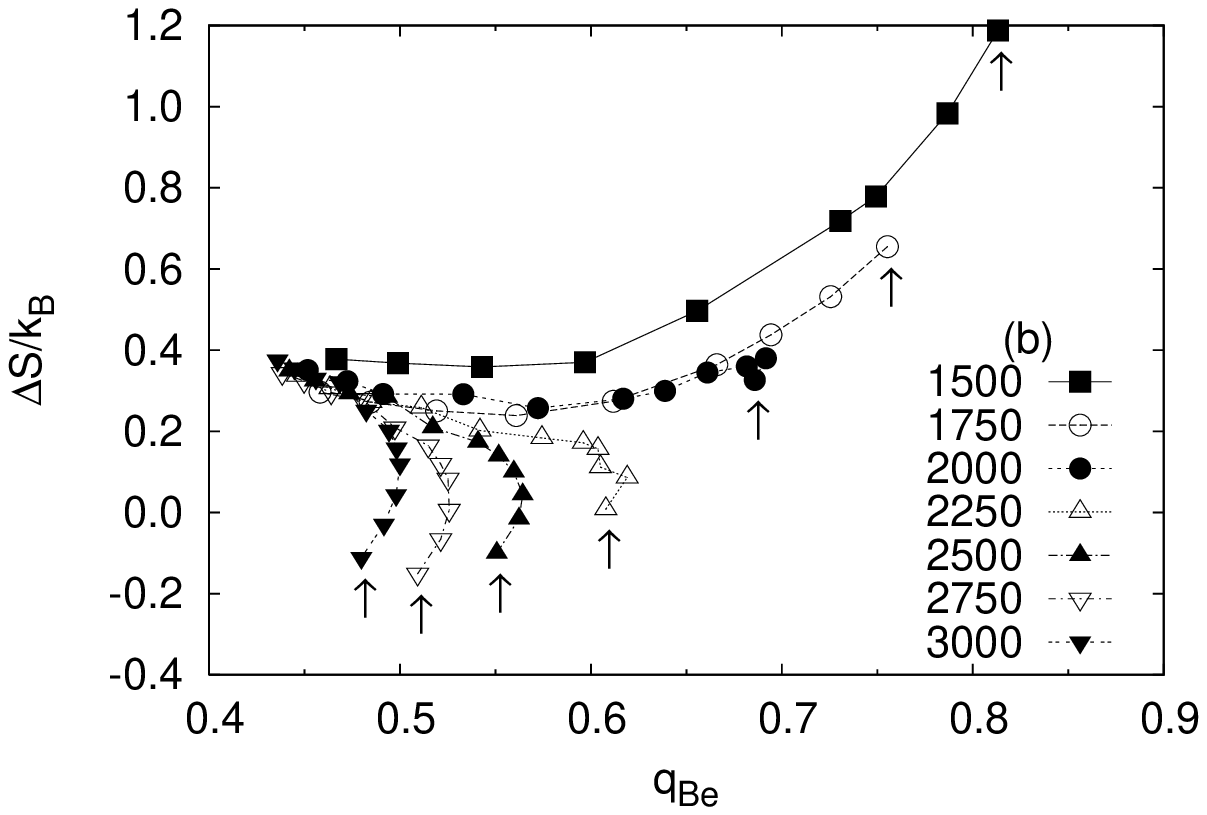}
\end{figure}

\newpage
\clearpage

\begin{figure} \caption{\quad} 
\includegraphics[scale=0.75]{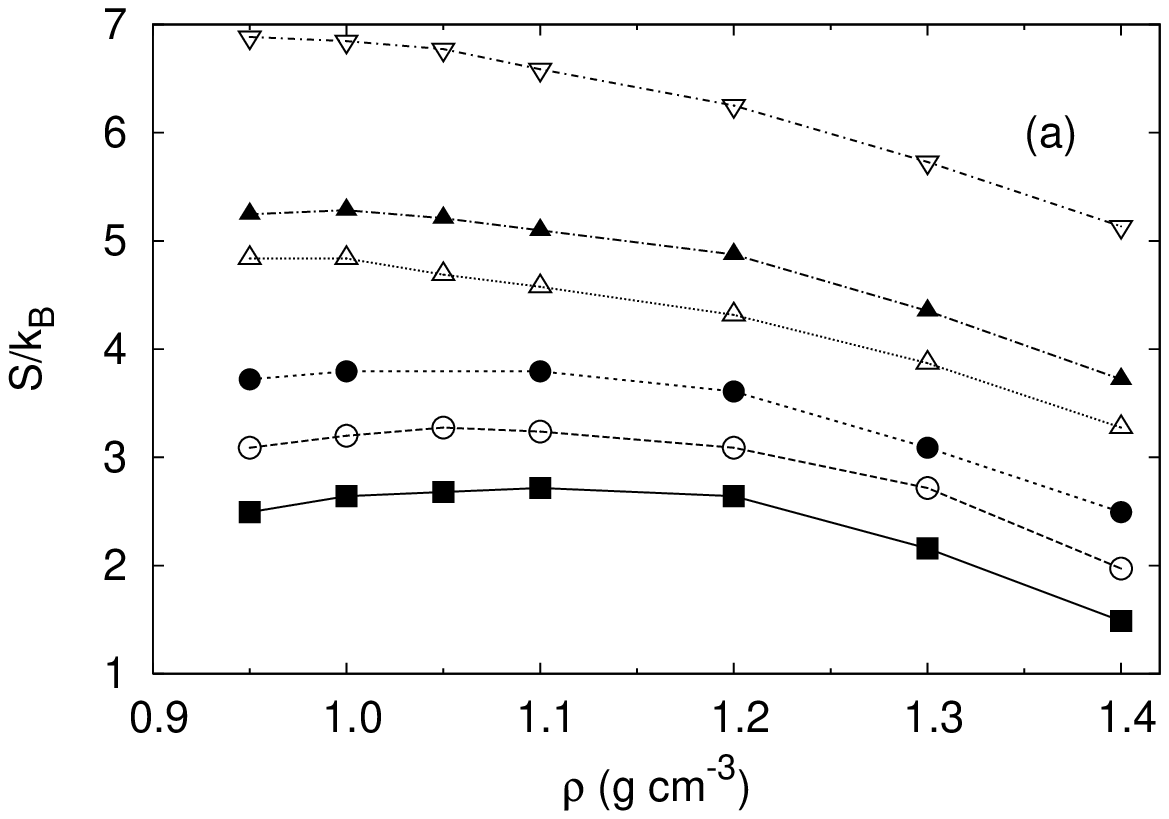}
\includegraphics[scale=0.75]{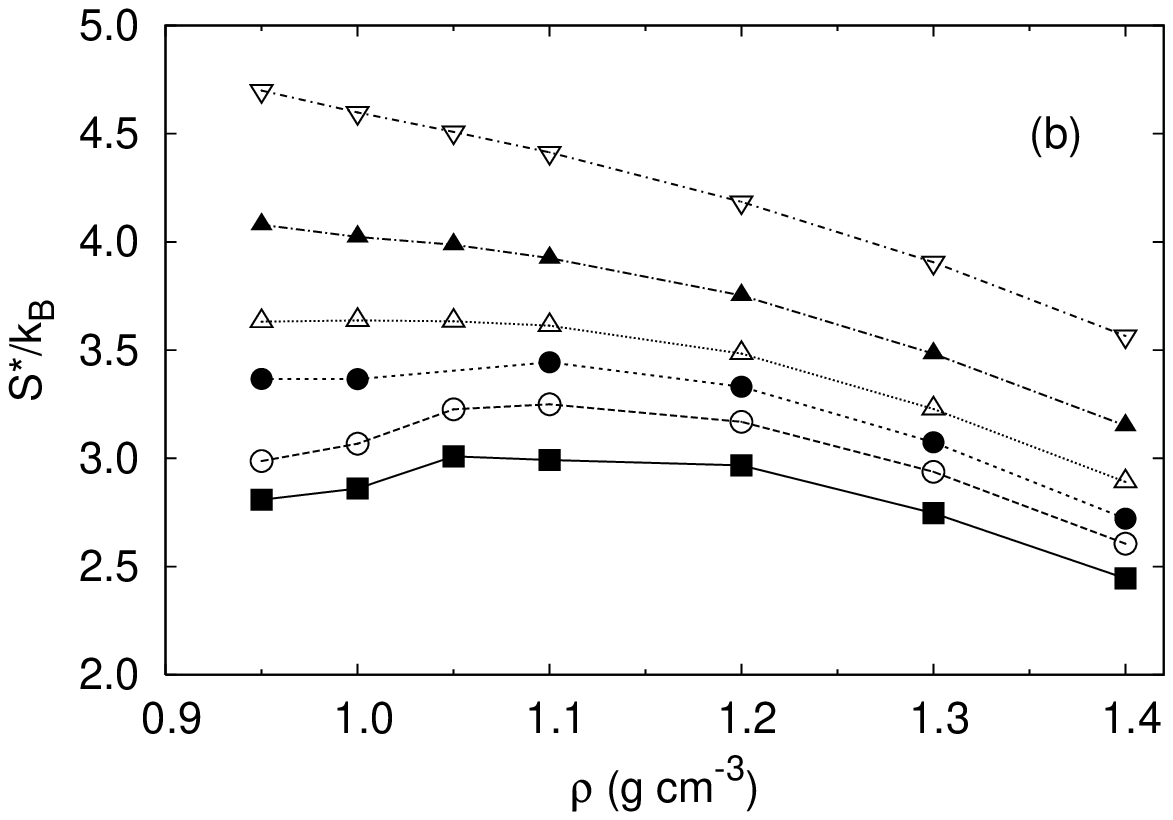}
\includegraphics[scale=0.75]{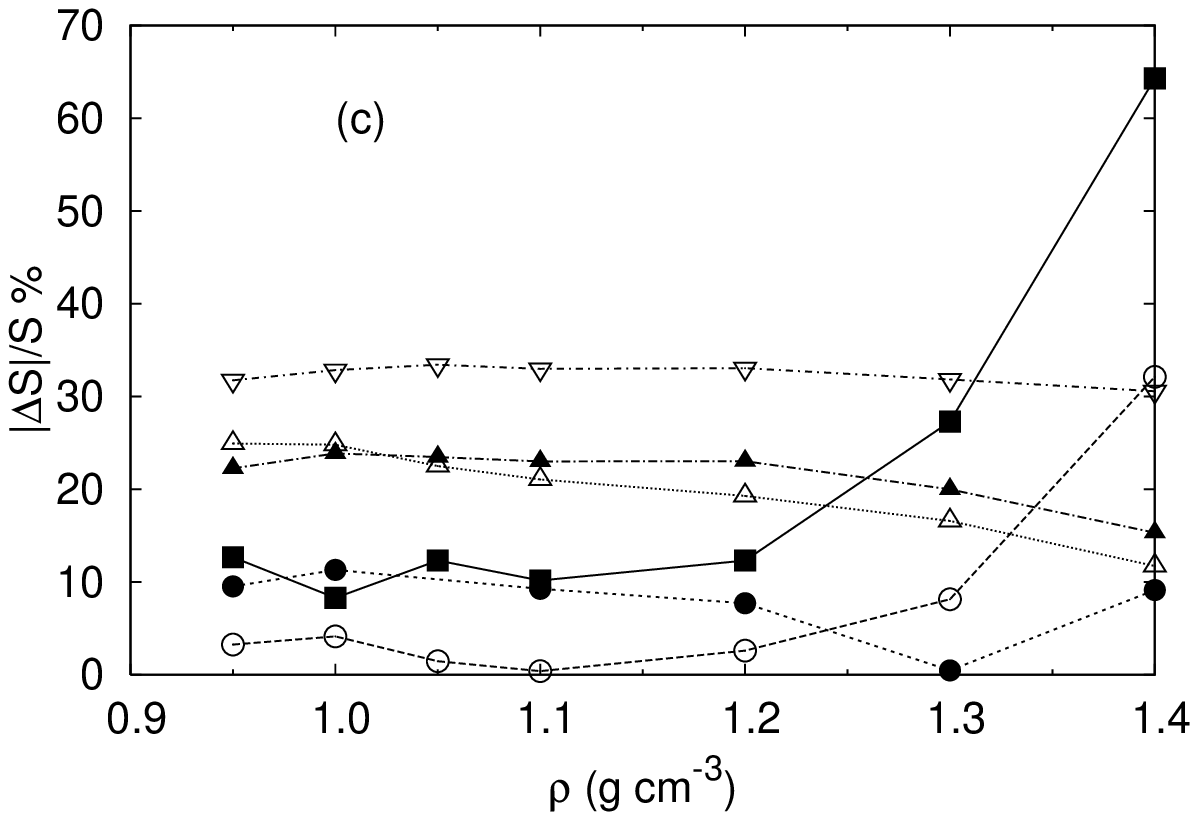}
\end{figure}

\newpage
\clearpage

\begin{figure} \caption{\quad}
\includegraphics{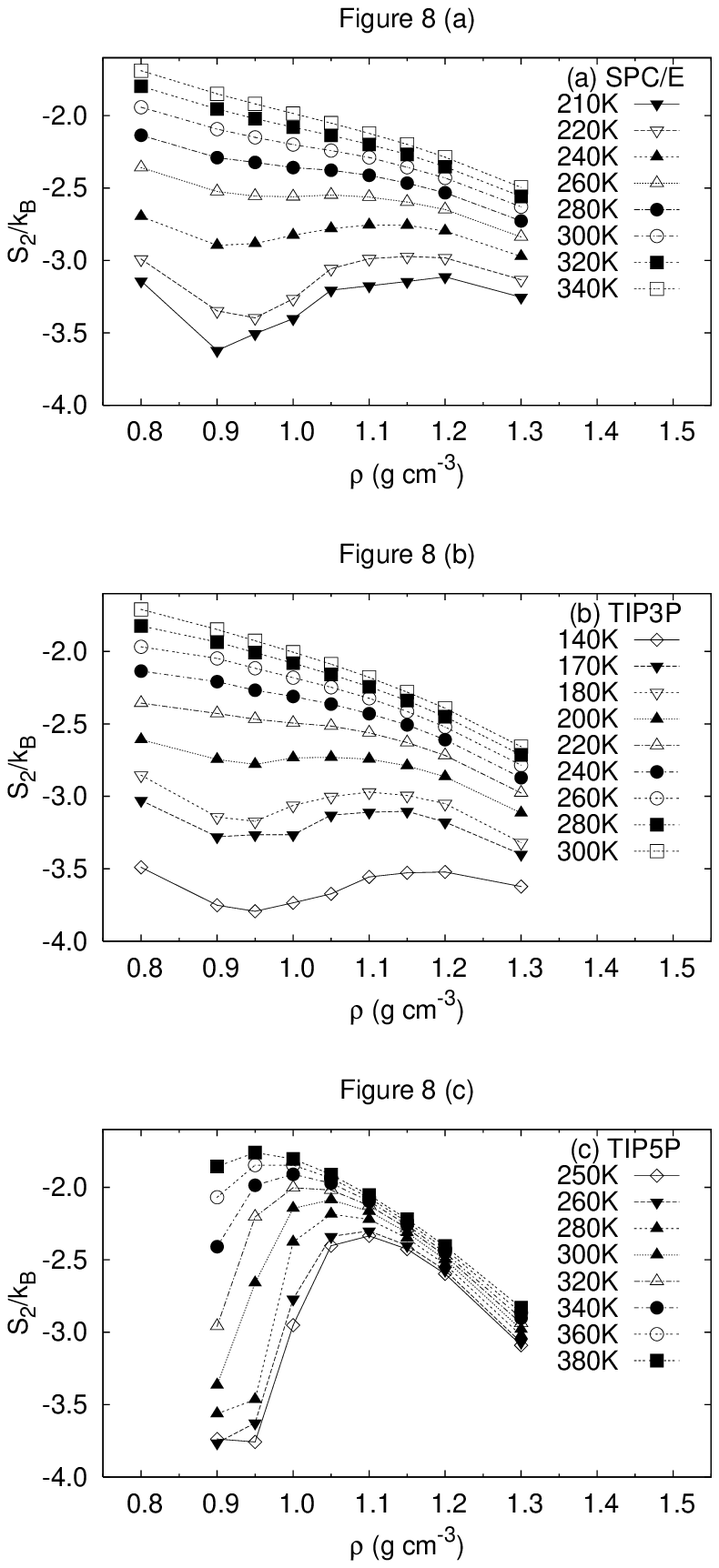}
\end{figure}

\end{document}